\documentclass[a4paper]{article} 

\usepackage{graphicx}
\usepackage{dcolumn}
\usepackage{bm}

\usepackage{amsmath,amssymb,amsthm} 

\begin{document}

\baselineskip=15pt
\newcommand{\RR}{\mathrm{I\!R\!}}
\newcommand{\II}{\mathrm{I\!I\!}}
\newcommand{\PP}{\mathrm{I\!P\!}}
\newcommand{\sn}{\mbox{sn}}
\newcommand{\am}{\mbox{am}}
\newcommand{\DEF}{\stackrel{\mbox{\rm\scriptsize def}}{=}}


\def\R{\hbox{\rm I\kern-.2150em R}}
\def\B{\hbox{\rm I\kern-.2150em B}}
\def\maps{:}
\def\into{\rightarrow}
\def\QED{\vrule height 6 pt width 5 pt depth 0pt}

\title{Writhe formulas and antipodal points in plectonemic DNA configurations}

\author{
S\'ebastien Neukirch \\
Institut Jean le Rond d'Alembert \\
CNRS \& UPMC Univ Paris 6 \\
4 place Jussieu (case 162) \\
75005 Paris, France,\\
and \\
Eugene L. Starostin \\
Centre for Nonlinear Dynamics, \\
Department of Civil, Environmental and Geomatic Engineering, \\
University College London, 
London, UK.}
\date{\today}
\maketitle 

\begin{abstract}
The linking and writhing numbers are key quantities when characterizing the structure of a piece of supercoiled DNA.
Defined as double integrals over the shape of the double-helix, these numbers are not always straightforward to compute, though a simplified formula exists \cite{fuller:1978}.
We examine the range of applicability of this widely-used simplified formula, and show that it cannot be employed for plectonemic DNA.
We show that inapplicability is due to a hypothesis of Fuller theorem \cite{fuller:1978} that is not met.
The hypothesis seems to have been overlooked in many works.
\end{abstract}
%
%
%

%
\\
\section{Introduction}
%
%
%
%
%
%
%
%
The double-helix structure of DNA is the source of many complications in its 
{\em in-vivo} functioning during condensation/decondensation, replication, or
transcription.
For example the necessary unzipping of the molecule during transcription induces torsional strain along DNA.
The development of magnetic tweezers
\cite{smith+al:1992,strick+al:1996}
or the use of quartz cylinders in optical tweezers 
\cite{Deufel:Nanofabricated-quartz-cylinders-for-angular-trapping:-DNA-supercoiling-torque-detection:2007}
allow researchers to investigate
the {\em in-vitro} response of DNA molecules to torsional stress.
Studies of the behaviour of this twist storing polymer are not just a game for physicists
as it has been clearly established that, e.g. the assembly of RecA could be
stalled by torsional constraints \cite{heijden+al:2005}, or that the rate of
formation and the stability of the complex formed by promoter DNA and RNA
polymerase depends on the torque present in the DNA molecule \cite{revyakin+al:2004}.
On the theoretical side, matters are made difficult by the nonlocality of
the topological property that is associated with the torsional constraint: the
link.
The two sugar-phosphate backbones of DNA have opposite orientation and the ends of a double-stranded DNA molecule can only be chemically
bound in such a way that each strand joins itself, thereby yielding two interwound closed curves (no M\"obius-like configuration can exist).
For a circularly closed DNA molecule, the link is the number of times one of
the sugar-phosphate backbone winds around the other.
Once the three dimensional shape of the molecule is projected onto a plane, the
link is given by half the number of signed crossings between the two
backbones.
This quantity is best seen as the number of turns put in an initially planar plasmid (a piece of circularly closed double-stranded DNA) before
closing it. Link has been shown to consist of two parts:
the two sugar-phosphate backbones of a plasmid can be linked because $(i)$ the
plasmid lies in a plane but the base-pairs are twisted around their centre line
(the curve joining the centroids of the base-pairs) and/or $(ii)$ the centre line
itself follows a writhed path in space.
In general the two possibilities coexist and the link $Lk$ of a DNA molecule
is the addition of the two quantities \cite{calugareanu:1959,calugareanu:1961,white:1969,Dennis:Geometry-of-Calugareanus-theorem:2005}:
\begin{equation}
Lk=Tw+Wr \, . \label{equa lk tw wr}
\end{equation}
The twist $Tw$ is a local quantity in the sense that it can be computed as the
single integral of the twist rate $\tau(s)$: $2 \pi \, Tw=\int_0^L \tau(s) \,
ds$,
where $s$ is the arclength along the centre line and $L$ the total contour length of the molecule.
In elastic DNA models, the twist rate $\tau(s)$ is normally coupled to mechanical quantities (e.g. the torque) characterizing the
molecule.
Contrary to the twist, Wr is a {\it global} property of the centre line of the
molecule. We first define the {\em directional writhe} of a closed curve $\Gamma$. Consider
a closed curve in 3D and project this curve, along a certain direction, on a plane. The number of signed crossings seen in the plane is the directional writhe {\em for that direction}. One could then reiterate the procedure with different directions of projection and compute the directional writhe for each direction. The average value obtained for the directional writhes, when all directions are considered, is the writhe $Wr(\Gamma)$ of the 3D curve $\Gamma$ \cite{fuller:1971}.
  

%
%
%
%
\section{Global and local writhe formulas} \label{section:global_and_local}
%
%
%
%
%
%
The following double integral has been introduced by C\u{a}lug\u{a}reanu \cite{calugareanu:1959,calugareanu:1961} and White \cite{white:1969}:
\begin{equation}
Wr^{CW}(\Gamma) = \frac{1}{4\pi} \int_\Gamma \int_\Gamma 
\frac{(\bm{t}(s) \times \bm{t}(s')) 
\cdot
(\bm{r}(s) - \bm{r}(s'))}{|\bm{r}(s)-\bm{r}(s')|^3} ds \, ds' \, ,
\label{equa:writhe_def}
\end{equation}
where $\bm{r}(s)$ is the position of points on $\Gamma$
and $\bm{t}(s)$ is the unit tangent to $\Gamma$.
As soon as a 3D curve does not self-intersect, 
the writhe $Wr(\Gamma)$ of the curve is given by Eq.~(\ref{equa:writhe_def}):
 $Wr(\Gamma)=Wr^{CW}(\Gamma)$ if $\bm{r}(s) \neq \bm{r}(s')$ $\forall s, s'$ with $s \neq s'$ \footnote{The double integral may be shown to converge for self-crossing curves provided the tangents at the intersection points are not aligned \cite{starostin:2002}.}.
The computation of the double integral of Eq. (\ref{equa:writhe_def}) is analytically hard and numerically time consuming to perform.
An important result though enables one to reduce the double integral to a
single integral, provided several hypotheses are fulfilled \cite{fuller:1978}.
It is the main purpose of this paper to show that these hypotheses are not met
in the case of plectonemic DNA.

Fuller's theorem \cite{fuller:1978} states that the writhe of the curve $\Gamma$ can be computed by
considering the writhe of a reference curve $\Gamma_0$ (which should be known
or easy to compute) and the continuous deformation (a homotopy) morphing $\Gamma_0$ to $\Gamma$:
\begin{eqnarray}
Wr(\Gamma) ~ {\stackrel{\mbox{\rm\scriptsize ?}}{=}} ~
 {Wr^F}(\Gamma,\Gamma_0) & \DEF & Wr(\Gamma_0) +  \frac{1}{2\pi} {Fu}(\Gamma,\Gamma_0) \label{equa:fuller_2nd_theo} \\
 \mbox{ where } {Fu}(\Gamma,\Gamma_0) &=&
 \int_0^{\sigma_L} 
\frac{\bm{t}_0(\sigma) \times \bm{t}(\sigma)}
{1+ \bm{t}_0(\sigma) \cdot \bm{t}(\sigma)} 
\cdot
\frac{d}{d\sigma} \left(\bm{t}_0(\sigma) + \bm{t}(\sigma)\right)
\, d\sigma \, ,
\nonumber
\end{eqnarray}
where $\bm{t}_0(\sigma) \DEF (d  \bm{r}_0(\sigma) / d \sigma)  / | d \bm{r}_0(\sigma) / d \sigma |$ is the unit tangent to $\Gamma_0$ and where $\sigma \in [0,\sigma_L]$ is a common parametrization for both curves, though not
necessarily the arclength along $\Gamma$ or $\Gamma_0$. The total contour length of $\Gamma$ is $L$.
The curve $\Gamma_0$ is normally chosen to be a fairly simple curve (e.g. planar with
zero writhe) so the computation of $Wr(\Gamma)$ boils down to estimating the
single integral in Eq.~(\ref{equa:fuller_2nd_theo}).
Before doing so, one should verify whether the hypotheses of Fuller's theorem
are fulfilled (this is the sense of the question mark in Eq.(\ref{equa:fuller_2nd_theo})).
The continuous deformation from $\Gamma_0$ to $\Gamma$ introduces
a familly of curves $\Gamma_\lambda$ with $\lambda \in [0,1]$ and $\Gamma_1 \equiv \Gamma$.
The unit tangent to $\Gamma_\lambda$
is  $\bm{t}_\lambda(\sigma) \DEF (d  \bm{r}_\lambda(\sigma) / d \sigma)  / | d \bm{r}_\lambda(\sigma) / d \sigma |$.
The first hypothesis is that none of the curves $\Gamma_\lambda$ self-intersects.
The second hypothesis is that, and this is the point we want to emphasize,
 there should be no point along any of the curves
$\Gamma_\lambda$ where $\bm{t}_\lambda(\sigma) \cdot \bm{t}_0(\sigma)=-1$.
For each value of $\lambda$, the (unit) tangent $\bm{t}_\lambda(\sigma)$ with $\sigma \in [0,\sigma_L]$ defines a curve, called the tangent indicatrix, on the unit sphere.
If for a certain $\bar{\lambda}$ and a certain $\bar{\sigma}$ we have $\bm{t}_{\bar{\lambda}}(\bar{\sigma}) \cdot \bm{t}_0(\bar{\sigma})=-1$ then, on the unit sphere, the point corresponding to $\bm{t}_{\bar{\lambda}}(\bar{\sigma})$  is antipodal to the point corresponding to $\bm{t}_0(\bar{\sigma})$.
For brevity we shall also call antipodal the point  $\bm{r}_{\bar{\lambda}}(\bar{\sigma})$ or the curve $\Gamma_{\bar{\lambda}}$ itself.
%

%

In models where self-intersection is prevented (e.g. models using hard- or soft-wall potentials of the molecule on itself) a continuous deformation free of self-intersection (an isotopy) may be easy to devise and the first hypothesis would be verified.
However nothing in these models ensure that the second hypothesis (that no antipodal points exist for all $\lambda \in [0,1]$ and all $\sigma \in [0,\sigma_L]$) is met and consequently Fuller's formula (\ref{equa:fuller_2nd_theo}) cannot be used to compute the writhe.
We nevertheless remark that even in the case of self-intersection or antipodal
points Fuller's formula is always correct {\em
  modulo} 2: the integer, and most important, part of $Wr(\Gamma)$ will not
be correct, but the fractional part will.
Each antipodal point present in the continuous deformation from $\Gamma_0$ to $\Gamma$ introduces a shift
of two units between the actual writhe $Wr(\Gamma)$ of the curve $\Gamma$ and the value $Wr^F(\Gamma,\Gamma_0)$ \cite{aldinger:1995}.
Consequently, if there are $m$ antipodal points
along the deformation $\lambda \in [0,1]$, the computed
value $Wr^F$ could be as far as $2m$ from the correct value $Wr$, but the fractional part will be accurate.

This discrepancy between $Wr^F$ and the actual writhe $Wr$ has already
been pinpointed in the case where DNA is treated as a fluctuating chain under low twist (i.e. without plectonemes) \cite{rossetto+maggs:2003,rossetto:2005,Heijden-Peletier-EtAl-On-end-rotation-for-open-rods-2007}.
Nevertheless we have found a certain number of references where 
Fuller's formula is used
\cite{fain+rudnick:1997,
fain+rudnick:1999,
moroz+nelson:1997,
bouchiat+mezard:1998,
bouchiat+mezard:2000,
moroz+nelson:1998,
nelson:1998,
marko:1998,
marko:1999,
bouchiat+mezard:2000,
zhou+al:2000,
lai+zhou:2003,
garrivier+fourcade:2000,
zhou+lai:2005,
haijun+al:1999,
zandi+rudnick:2001,
zhou+lai:2003,
zhou:2005,
sinha:2004}.
In most cases the hypotheses of Fuller's theorem were not checked, if only mentioned.
In some works the formula is used in a scheme that provides
an estimate for the torsional stiffness of the DNA molecule
\cite{bouchiat+mezard:1998,bouchiat+mezard:2000} and this has been shown to lead to incorrect results \cite{rossetto+maggs:2002,rossetto:2005,rossetto+maggs:2003}.
Nevertheless, in some few papers \cite{sinha:2004,moroz+nelson:1997,moroz+nelson:1998,nelson:1998}, the formula was used to asses the writhe of DNA configurations under high stretching force and low twist, in which cases antipodal points are absent and the formula is correct. 
%

%

%
%
%
%
%
%
%

%
%
%
%
%
\section{The writhe and link of DNA molecules in magnetic tweezer experiments}
\begin{figure}[htbp]
\begin{center}
\includegraphics[angle=0,width=0.40\linewidth]{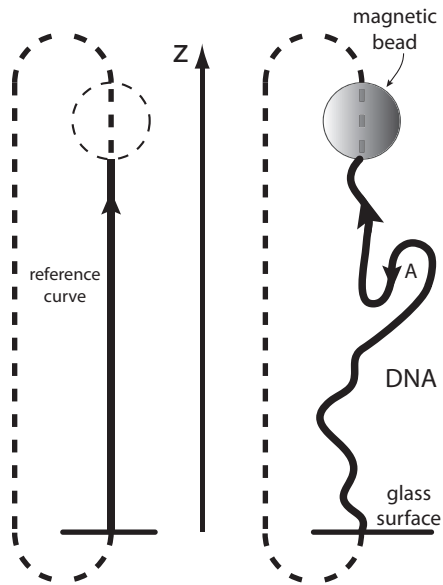}
\caption{The shape of a DNA molecule subjected to a vertical pulling force follows a 3D curve  which is locally made random by thermal agitation. As explain in \cite{rossetto+maggs:2003}, when the force is small or moderate, points where the tangent to the molecule is oriented downward can appear, e.g. point A. These points on the curve are antipodal to corresponding points on the reference curve.}
\label{fig:antipodal_WLC}
\end{center}
\end{figure}

%
%
%
%
%
During a force-extension experiment on a single DNA molecule thermal agitation deforms the molecule
whose shape locally adopts random directions (Fig.~\ref{fig:antipodal_WLC}, left).
In the absence of (or under low) twist, 
DNA is modeled  as a worm-like chain
\cite{kratky+porod:1949}; the path followed by its centre line in space looks
like a (directed) random walk \cite{moroz+nelson:1997}.
In such configurations,
the writhe is usually evaluated using Fuller's formula (\ref{equa:fuller_2nd_theo}) with the reference curve shown in Fig.~\ref{fig:antipodal_WLC}, left.
Since the writhe is classically defined for a closed curve, the
reference and actual curves are all closed by imaginary $C$-shaped curves (dashed in  Fig.~\ref{fig:antipodal_WLC})
that connect the top to the base of the configurations.
(Other choices of closures and interferences due to the closure in the computation of the writhe are discussed in \cite{rossetto+maggs:2003,starostin:2002,starostin:2005,Starostin:Comment-on-Cyclic-rotations-contractibility-and-Gauss-Bonnet:2002,Heijden-Peletier-EtAl-On-end-rotation-for-open-rods-2007}.)
Strictly speaking the curve $\Gamma$ for which we compute the writhe consists of two parts: the closure and the part corresponding to the molecule.
As the closure remains unchanged in the continuous deformation $\lambda \in [0,1]$ its contribution to Fuller's integral is zero.
Consequently we will forget about the closure and call $\Gamma$ the (open) curve described by the molecule only.
The reference curve $\Gamma_0$ is then the $z$ axis, and Fuller's formula (\ref{equa:fuller_2nd_theo}) becomes:
\begin{eqnarray}
Wr^F(\Gamma,\Gamma_0)&=&Wr(\Gamma_0)+ \frac{1}{2 \pi} \int_0^L 
\frac{\bm{e_z} \times \bm{t}(s) }{1+ \bm{e_z} \cdot \bm{t}(s)}
\cdot 
\frac{d }{ds}\bm{t}(s)  \, ds \nonumber \\
&=&0+\frac{1}{2 \pi}\int_0^L 
 \left(1-\cos \theta(s)\right) \, \frac{d\psi(s)}{ds} \,ds \, ,
\label{equa:fuller_euler_angle}
\end{eqnarray}
if the tangent $\bm{t}(s)=(\sin \theta \cos \psi , \sin \theta \sin \psi, \cos
 \theta)$ is parametrized by Euler angles $\theta(s) \in [0,\pi]$, and $\psi(s) \in [0, 2\pi)$.
As stated above, Eq.~(\ref{equa:fuller_euler_angle}) is only valid if the curve $\Gamma$ can be deformed into the reference curve $\Gamma_0$ without
passing through configurations $\Gamma_\lambda$ having their tangent vector $\bm{t}_\lambda(\sigma)$ facing -$\bm{e_z}$: $\bm{t}_\lambda(\sigma) \neq -\bm{e_z}$ for all $\lambda$, $\sigma$. We remark that in the present case where
$\Gamma_0$ is the $z$ axis an antipodal point is a point where the curve $\bm{t}_\lambda(\sigma)$ passes through the south pole of the unit sphere: $\theta(\sigma)=\pi$.
\begin{figure}[htbp]
\begin{center}
\includegraphics[angle=0,width=0.4\linewidth]{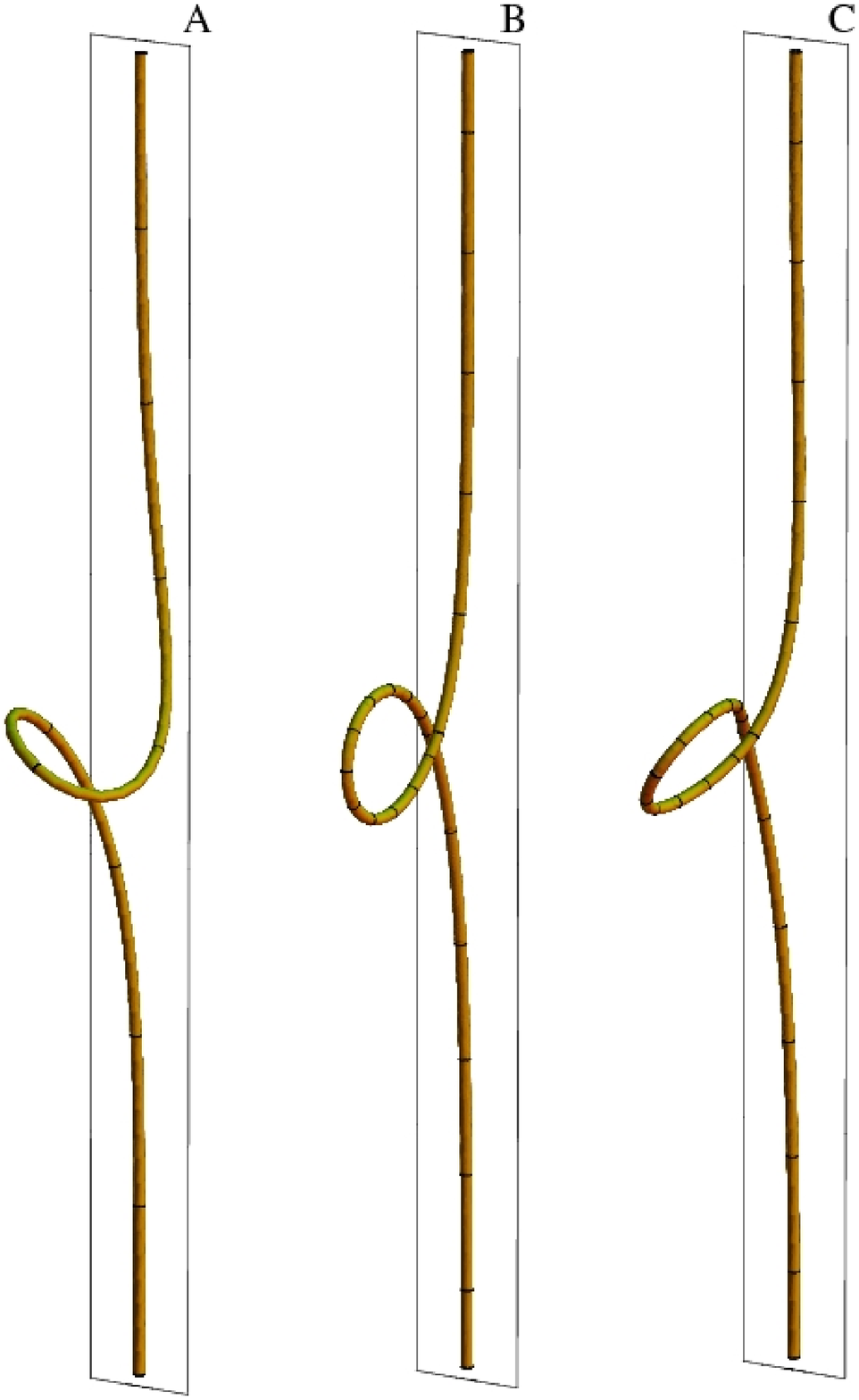} 
\includegraphics[angle=0,width=0.4\linewidth]{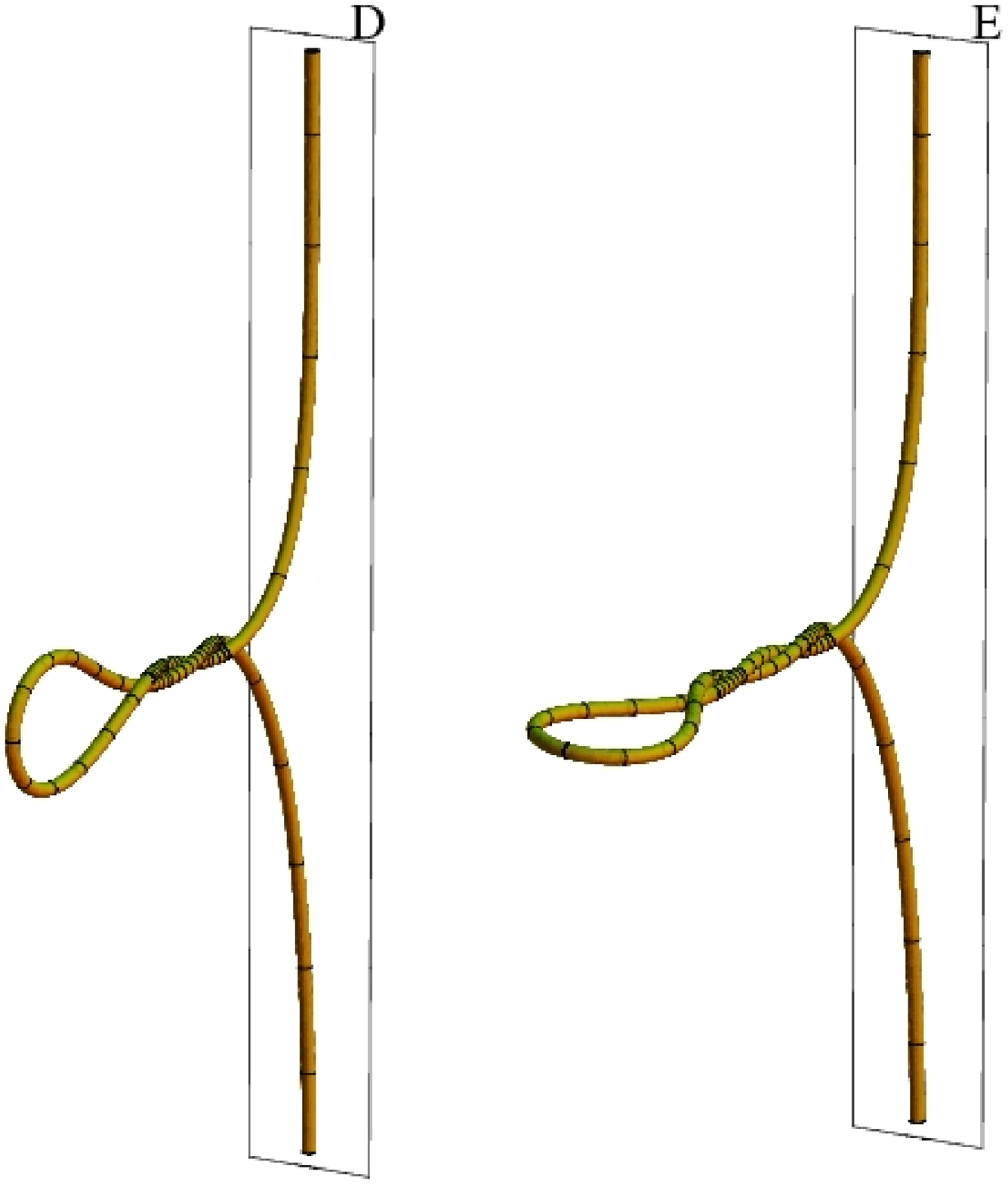} 
\caption{Configurations $A$, $B$, $C$, $D$, and $E$ (see Fig.~\ref{fig:diag_z_de_n}) obtained by numerical continuation \cite{neukirch:2004}. Configurations $B$ and $D$ each have an antipodal point to the reference curve of Fig.~\ref{fig:antipodal_WLC}. These antipodal points are located at the middle point of the end-loop of the plectonemic structure.}
\label{fig:antipodal_plectonems}
\end{center}
\end{figure}

In the case where the actual configuration $\Gamma$ exhibits a single or multiple antipodal points  it has been proposed in \cite{bouchiat+mezard:2000} to evaluate the writhe from Eq.~(\ref{equa:fuller_euler_angle}) with Euler angles defined on a truncated unit sphere: $\theta$ would not be allowed to reach $\pi$ and hence antipodal points would be avoided (the curve $\tilde{\Gamma}$ so defined would be very near the real curve $\Gamma$, hence the writhes would almost be the same).
We stress that this process is not sufficient as even for a given curve with no antipodal points, it is not clear whether Eq.~(\ref{equa:fuller_euler_angle})
is valid or not. In order for Eq.~(\ref{equa:fuller_euler_angle})
to be valid, one has to exhibit a continuous deformation from the $z$ axis to the curve $\Gamma$ that is entirely free of antipodal points.
There are many cases where
the actual curve $\Gamma$ is free of antipodal points, but where
a deformation free of antipodal points does not exist.
Consequently in these cases Eq.~(\ref{equa:fuller_euler_angle}) yields an incorrect result (unless antipodal points of opposite signs cancel out).
Nevertheless, it has been noted that under high stretching force (e.g. $F \sim
5$ pN) and low torque the DNA molecule is almost straight and no such antipodal points exist \cite{nelson:1998}.
In such a case Fuller's formula is correct provided no plectonemes are present, as we will see now.

\begin{table}[ht]
\begin{center}
\begin{tabular}{| l | c | c | c | c | c |}
\hline
  & ~A~ & ~B~  & ~C~  & ~D~  & ~E~ \\
\hline
$Wr$ & 0.70 & 0.99 & 1.09 & 2.92  & 3.45   \\
\hline
$Wr^F$  & 0.70 & - & -0.91 & - & -0.55 \\
\hline
$Lk$ & 4.46 & 1.84 &  4.15 &  4.23  &  4.75   \\
\hline
$Lk^F$  &  4.46 & - &  2.15 & - &  0.75 \\
\hline
\end{tabular}
\caption{Writhe (first line) and Link (third line) of the configurations of Fig.~\ref{fig:antipodal_plectonems}, computed by continuation or use of the double integral (Eq.~(\ref{equa:writhe_def})).
The second line is computed from Eq.~(\ref{equa:fuller_euler_angle}) and the last line is computed from Eq.~(\ref{equa:link_phi_psi}).
Formulas  (\ref{equa:fuller_euler_angle}) and (\ref{equa:link_phi_psi}) are not applicable
on configurations $B$ and $D$ that each have an antipodal point, and yield incorrect results for configurations $C$ and $E$.
}
\label{table:wr_lk_wrf_lkf}
\end{center}
\end{table}

\begin{figure}[htbp]
\begin{center}
\includegraphics[angle=0,width=0.9\linewidth]{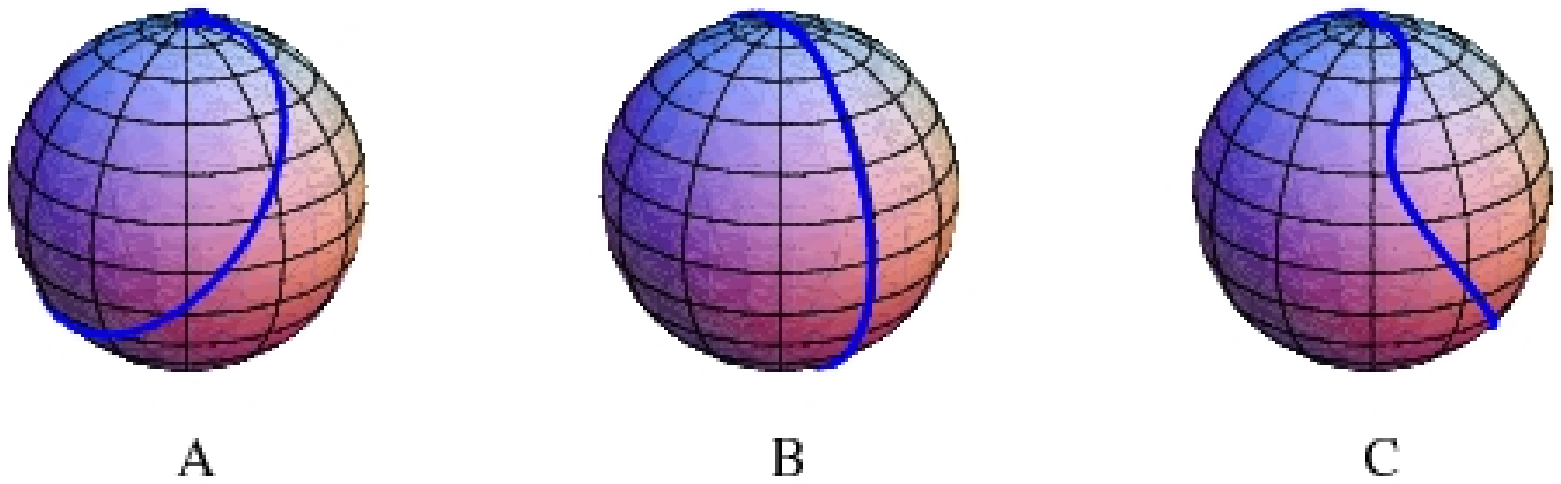} \\
\includegraphics[angle=0,width=0.9\linewidth]{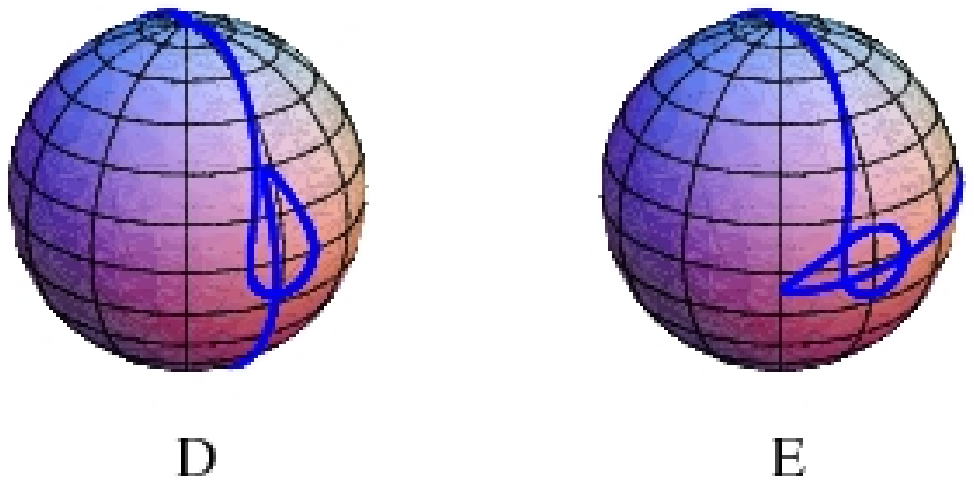} 
\caption{Tangent indicatrices of configurations $A$, $B$, $C$, $D$, and $E$ of Fig.~\ref{fig:antipodal_plectonems}. An antipodal point arises when the curve passes through the south pole of the sphere, i.e. for configurations $B$ and $D$.}
\label{fig:unit_spheres}
\end{center}
\end{figure}
%

%

In magnetic tweezer experiments, when a large amount of turns are put in (by
rotation of the magnetic bead around the $z$ axis), the DNA molecule
reacts by forming plectonemes \cite{strick+al:1998}.
The number of turns $n$ imposed on the magnetic bead is given by the Link $Lk$ of the molecule.
We now show that the presence of plectonemes in the supercoiled configuration
prevents the existence of a deformation, from $\Gamma_0$ (i.e. the $z$ axis) to $\Gamma$, that is free of antipodal points, and consequently forbid the use of  Eq.(\ref{equa:fuller_euler_angle}).

\begin{figure}[htbp]
\begin{center}
\includegraphics[angle=0,width=0.70\linewidth]{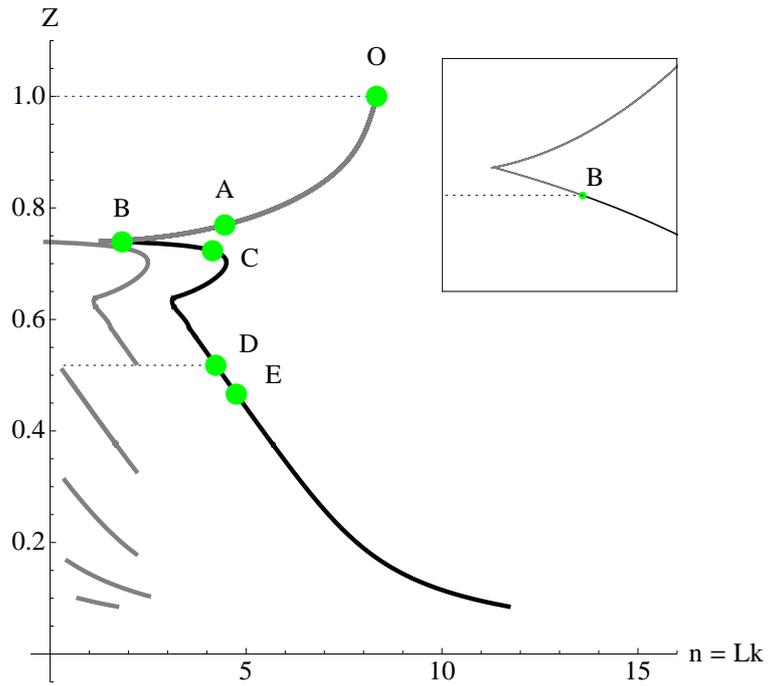}
\caption{The dark continuous curve is obtained from numerical continuation \cite{neukirch:2004} and the broken gray curve is obtained from Eq.~(\ref{equa:link_phi_psi}).
Five configurations $A$, $B$, $C$, $D$, and $E$ are selected and plotted in Fig.~\ref{fig:antipodal_plectonems} together with their tangent indicatrices in Fig.~\ref{fig:unit_spheres}.
The inset shows a zoom around point $B$, the first antipodal point, where the gray curve jumps by an amount of -2 units.
From $O$ to $B$ the two curves coincide.}
\label{fig:diag_z_de_n}
\end{center}
\end{figure}

In terms of the Euler angles the twist of the molecule can be computed as
\begin{equation}
Tw=\frac{1}{2\pi} \int_0^L \left( \frac{d\phi}{ds} + \cos\theta \, \frac{d\psi}{ds} \right) \, ds,
\label{equa:twist}
\end{equation}
where $\phi(s)$
is the third Euler angle (see e.g. \cite{heijden+thompson:2000}).
Using Eqs.~(\ref{equa:fuller_euler_angle}) and (\ref{equa:twist}) we obtain:
\begin{equation}
Lk^F=\frac{1}{2\pi} \int_0^L \frac{d}{ds} (\phi + \psi) \, ds \, .
\label{equa:link_phi_psi}
\end{equation}
This formula usually is the starting point for computations of the link of supercoiled  configurations, see e.g. \cite{bouchiat+mezard:2000} or \cite{fain+rudnick:1997}.
We show here that it yields incorrect results when plectonemes are present.
We have performed computations to model the elastic response of a twist
storing filament subject to tensile and
torsional constraints and we quantitatively reproduced the plectonemic regime
characterised by the linear decrease of the end-to-end distance of the filament
as a function of the number $n$ of turns put in \cite{neukirch:2004}.
The plectonemic configurations were computed numerically using a continuation
algorithm and $n=Lk$ was computed by continuity so that no integer number of
turns is missed (we also performed numeric integration of the double integral of Eq.~(\ref{equa:writhe_def}) for the writhe of the configurations, with closures, and always obtained consistent results).
These shapes serve as an illustration for the computation of the writhe and consequently are used for their geometry only. The fact that they are mechanical equilibria is not relevant here.
The continuous dark curve in Fig.~\ref{fig:diag_z_de_n} shows an output of the numerics, drawn in the $(Lk,Z)$ plane, where $Z$ is the vertical extension of the molecule.
The curve starts at point $O$ which corresponds to a straight and twisted configuration.
The path is then monotonically decreasing in $Z$. We have selected five configurations $A$, $B$, $C$, $D$, and $E$ which are drawn in Fig.~\ref{fig:antipodal_plectonems}, together with their corresponding tangent indicatrices (see definition in Section \ref{section:global_and_local}) drawn in Fig.~\ref{fig:unit_spheres}.
Configurations $B$ and $D$ each comprise an antipodal point located at the middle point of the end-loop of the plectonemic structure.
This can be verified in Fig.~\ref{fig:unit_spheres}-$B$ and $D$ where the tangent indicatrices  pass through the south pole of the unit sphere.
Using the same geometric configurations, we compute $Lk^F$ from Eq.~(\ref{equa:link_phi_psi}) and we plot the corresponding curve, in gray, on the same diagram.
%
We see that each antipodal event introduces a shift of two units in the gray curve, which is consequently broken.
As expected this confirms that Eq.~(\ref{equa:link_phi_psi}) is only valid {\em modulo} 2 and hence should not be used to estimate the link of plectonemic configurations.
The writhe and link of the five example configurations are given in Table~\ref{table:wr_lk_wrf_lkf} where they are compared with $Wr^F$ and $Lk^F$ given by Eqs.~(\ref{equa:fuller_euler_angle}) and (\ref{equa:link_phi_psi}).
For configuration $E$, which is separated from the reference curve by two antipodal events, we see that $Wr^F$ (resp. $Lk^F$) is 4 units away from the correct $Wr$ (resp. $Lk$) value.
This configuration $E$ is an illustration of the fact that Fuller's formulas (Eq.~(\ref{equa:fuller_euler_angle}) or (\ref{equa:link_phi_psi})) can be wrong even for a configuration that does not comprise any antipodal point, which is clearly illustrated in Fig.~\ref{fig:unit_spheres}-$E$ where we see that the tangent indicatrix is nowhere near the south pole.



We first comment on the size of the gap between the two curves. A plectonemic DNA configuration $\Gamma$ that shows $N_x$ (positive) crossings on a lateral projection has $Wr(\Gamma)\simeq +N_x$ \cite{Thompson:Cutting-DNA:-mechanics-of-the-topoisomerase:2008}.
If we continuously deform this configuration to the reference curve $\Gamma_0$ of Fig.~\ref{fig:antipodal_WLC} left by unwinding the plectonemic region, an antipodal point arises each time the tangent at the apex of the terminal loop (at the end of the plectonemic region) is facing downward.
This happen $N_x / 2$ times.
As proved in \cite{aldinger:1995} the presence of $m=N_x / 2$ antipodal points
leads to a discrepancy in Fuller's formula of $2m$: $| Wr - Wr^F | \leq 2m$.
Since $2m=N_x \simeq Wr$, we have that: $0 \lesssim Wr^F \lesssim 2Wr$, which corresponds to an error of up to 100\%. This is apparent in Fig.~\ref{fig:diag_z_de_n} where the broken gray curve stays near the vertical axis while the (continuous) dark curve monotonically increases in Link.

Second we note that the discrepancy between the two curves
%
%
occurs shortly after self-contact has started in the filament, for $n \simeq 1$, see inset of Fig.~\ref{fig:diag_z_de_n}.
Plectonemic structure may exist for small number of turns $n$, provided the pulling force is not too large.
On the other hand, a large pulling force does not rule out the occurence of plectonemes, provided that $n$ is large enough.
This leaves a small parameter regime (large pulling force, low number of turns) where plectonemes are absent.
When the two sources of discrepancies are considered (random walk antipodal
points \cite{rossetto+maggs:2003,rossetto:2005}, and plectonemic antipodal points) one sees that the use of
Eq.~(\ref{equa:fuller_euler_angle}) (resp. Eq.~(\ref{equa:link_phi_psi}))
to compute the writhe (resp. the link) in a model for DNA under tensile
and/or torsional stress is to be avoided unless $n \simeq 0$ and the tensile force is
large. 
%

%
%
%
%
%
%
%
%
%
%
\section{Discussion and concluding remarks} \label{section conclusion}
%
%
%
%
%
%
%




We summarize here few properties of the quantities $Wr^{CW}(\Gamma)$, $Wr^{F}(\Gamma,\Gamma_0)$ which do not always give the writhe of a curve $\Gamma$.
The quantity $Wr^{CW}(\Gamma)$ is a function of the curve $\Gamma$ only, whereas the quantity $Wr^{F}(\Gamma,\Gamma_0)$ also depends on the reference curve $\Gamma_0$. 
The quantity $Wr^{CW}$ yields the correct value for the writhe $Wr$ of a closed curve as soon as the curve is not self-intersecting.
Along a continuous deformation $\Gamma_\lambda$ with $\lambda \in [0,1]$ (with $\Gamma_1 \equiv \Gamma$), the quantity $Wr^{CW}(\Gamma_\lambda)$ jumps by two units when the curve $\Gamma_\lambda$ intersects itself (say at $\lambda=1/2$). Since the writhe $Wr$ also jumps by two units,  the quantity $Wr^{CW}$ is equal to $Wr$ before and after the self-crossing event, i.e. $Wr^{CW}(\Gamma_\lambda)=Wr(\Gamma_\lambda)$, $\forall \lambda \neq 1/2$.
Now the quantity $Wr^{F}$ has no such discontinuity:  along a continuous deformation where the curve self-intersects, the writhe $Wr$ will jump by two units, but the quantity $Wr^{F}$ will stay continuous. This means that the quantity $Wr^{F}$ no longer yields the correct value for the writhe {\em after} the self-crossing event,  $Wr^{F}(\Gamma_\lambda,\Gamma_0) = Wr(\Gamma_\lambda)$, $\forall \lambda < 1/2$ but $Wr^{F}(\Gamma_\lambda,\Gamma_0) \neq Wr(\Gamma_\lambda)$, $\forall \lambda > 1/2$.
The same is true for antipodal points.
Along a continuous deformation $\Gamma_\lambda$ with $\lambda \in [0,1]$, the quantity $Wr^{F}(\Gamma_\lambda,\Gamma_0)$ jumps by two units when the curve $\Gamma_\lambda$ has an antipodal point (say at $\lambda=1/2$) with regard to the reference curve $\Gamma_0$. On the other hand the writhe $Wr$ stays continuous. This means that the quantity $Wr^{F}$ no longer yields the correct value for the writhe as soon as an antipodal event happens,  $Wr^{F}(\Gamma_\lambda,\Gamma_0) = Wr(\Gamma_\lambda)$, $\forall \lambda < 1/2$ but $Wr^{F}(\Gamma_\lambda,\Gamma_0) \neq Wr(\Gamma_\lambda)$, $\forall \lambda > 1/2$.
This is an important point and many authors seem to believe that the quantity $Wr^{F}(\Gamma_\lambda,\Gamma_0)$ only has problems for configurations actually comprising an antipodal point ($\lambda=1/2$ in the above example).
Therefore  the quantity $Wr^{F}(\Gamma,\Gamma_0)$ may not be equal to the writhe $Wr(\Gamma)$ even for curves that {\em do not} comprise any antipodal point. This make the use of $Wr^{F}(\Gamma,\Gamma_0)$ uneasy, as one has to first verify the absence of antipodal events in the {\em entire} continuous deformation $\Gamma_0 \to \Gamma$ .
On the contrary the use of
$Wr^{CW}(\Gamma)$ is much easier in the sense that one just has to check that the actual curve $\Gamma$ does not self-intersect.
In this sense the quantities $Wr^{CW}(\Gamma)$ and $Wr^{F}(\Gamma,\Gamma_0)$ do not suffer from the same pathologies in the computation of the writhe, contrarily to what is claimed in \cite{samuel+al:2006}.
Another consequence is that, when sampling DNA configurations to construct a statistical ensemble and compute writhe averages and fluctuations, it is not enough to introduce, as was done in \cite{bouchiat+mezard:2000}, a small forbidden region around the south pole of the unit sphere to ensure that $Wr^{F}(\Gamma,\Gamma_0)$ yields a correct value.
In fact many of these sampled configurations, even free of antipodal points, are configurations that suffer the same problems as the configurations with $\lambda>1/2$ above: $Wr^{F}(\Gamma,\Gamma_0) \neq Wr(\Gamma)$, as numerically verified in \cite{rossetto+maggs:2003,rossetto:2005}.

Finally we want to point out the following property. We saw that if a continuous deformation $\Gamma_\lambda$ with $\lambda \in [0,1]$ contains an antipodal event, then $Wr^{F}(\Gamma,\Gamma_0)$ does not yield the correct result after the antipodal event.
Now one can argue that yet another continuous deformation, free of antipodal events and self-crossings, may exist (with the same initial and final curves) and that in this case $Wr^{F}(\Gamma,\Gamma_0)$ would yield the correct result. This is not the case since, as we show in Appendix B, as soon as a corrupted deformation exists in between two curves $\Gamma_0$ and $\Gamma$, then {\em all} continuous deformations are corrupted and $Wr^{F}(\Gamma,\Gamma_0)$ definitely yields an incorrect result.
This means that the reference curve $\Gamma_0$ cannot
be used. One is bound to find another reference curve or to use the double integral $Wr^{CW}$.
It appears that an easy way to verify the applicability of Fuller formula (\ref{equa:fuller_2nd_theo}) between an actual curve and a reference curve is to consider any convenient deformation that avoids antipodal events.
This can always be done if one allows for self-crossings.
Then formula  (\ref{equa:fuller_2nd_theo}) is applicable if and only if the net sum of self-crossings is zero. 

In the case of numerical computations, either dealing with continuation of mechanical equilibria or statistical ensembles of configurations, the writhe can be assessed  in an accurate way by using both the double integral (\ref{equa:writhe_def}) and Fuller integral (\ref{equa:fuller_euler_angle}) in a cooperative way.
The double integral being time consuming to evaluate, one can in a first step discretize it \cite{swigon+al:1998} with a reduced number of elements so that it produces an approximate result that only has to be accurate up to $\pm 1$ (one still has to estimate how many elements are needed to obtain such an accuracy \cite{cantarella:2002}). 
Then in a second step, Fuller integral is used to refine the result. Even if Fuller integral may be off by several integers, its fractional part is correct.
In this scheme, the double integral yields the integer part of the writhe and Fuller integral yields the fractional part of the writhe.
Moreover in cases where one knows the correct value of the writhe of a nearby configuration (e.g. a predecessor configuration, one move away in a Monte-Carlo scheme) this value can be used as the approximate result of the first step and one only has to compute Fuller integral to obtain a correct and accurate value of the writhe. (The usual assumption that there is no self-crossing between the predecessor and the actual configuration still holds.)

In conclusion we have shown, by producing counter-examples and explaining the underlying causes, that formulas (\ref{equa:fuller_euler_angle}) and (\ref{equa:link_phi_psi}) cannot be used to compute the writhe and the link of supercoiled DNA configurations encountered in magnetic tweezer experiments.

\section{Acknowledgments}
The work of E.S. on this article was supported by the UK's Engineering and Physical Sciences Research Council under grant number GR/T22926/01.

%
%
%
%
%
%
%
%
%
\section{Appendix A : Fuller's formula in a detailed example} \label{appendix 1}
\begin{figure}[htbp]
\begin{center}
\includegraphics[angle=0,width=0.70\linewidth]{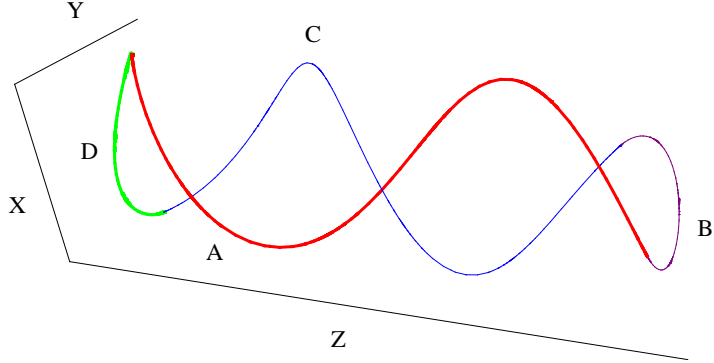}
\caption{Circularly closed supercoiled plasmid corresponding to curve $\Gamma = \Gamma_A \cup \Gamma_B \cup \Gamma_C \cup \Gamma_D$ with $\theta_A=\pi/6$ and $L_A=5\pi$.}
\label{fig:clamped_ply}
\end{center}
\end{figure}

%
%
%
%
%
%
%
%
%
In this appendix we compare, for a given example curve, the value computed from the double integral $Wr^{CW}$ to the value computed from the single integral $Wr^F$.
We start by introducing the parametrization of a circularly closed supercoiled plasmid (see Fig.~\ref{fig:clamped_ply}).
The parametrization is divided in four parts, called $A$, $B$, $C$, and $D$. Part $A$ is a right-handed helix of radius unity, total arclength $L_A$, and helical angle $\theta_A$ (with $0 \leq \theta_A < \pi/2$):
\begin{equation}
\bm{r_A}= \left(
\begin{array}{c}
 \sin \psi_A(s_A)\\
- \cos \psi_A(s_A) \\
s_A \, \cos \theta_A
\end{array}
\right)
\; \mbox{ and } \;
\bm{t_A}=\frac {d \bm{r_A}}{d s_A}  = \left(
\begin{array}{c}
\sin {\theta_A} \, \cos \psi_A(s_A)\\
\sin \theta_A \, \sin \psi_A(s_A) \\
\cos \theta_A
\end{array}
\right)
\end{equation}
with $s_A \in [0,L_A]$ and $\psi_A(s_A)=s_A \, \sin \theta_A$. We note $\Delta \psi_A \DEF L_A \, \sin \theta_A$. We have $| \bm{t_A}(s_A) |=1$ for all $s_A$, and consequently $s_A$ is the arc length in part $A$. 
The end loop $B$ connects the two helices $A$ and $C$:
\begin{equation}
\bm{r_B}= \left(
\begin{array}{c}
 \sin \Delta \psi_A \, \cos s_B + \cos \Delta \psi_A \sin \theta_A \sin s_B \cos s_B  \\
 -\cos \Delta \psi_A \, \cos s_B + \sin \Delta \psi_A \sin \theta_A \sin s_B \cos s_B \\
L_A  \, \cos \theta_A +  \left( \sin \theta_A - \sin \theta_B(s_B)  \right) / (1 + 2 \, \theta_A/\pi) 
\end{array}
\right)
 \mbox{ and }  \bm{t_B}=\frac {d \bm{r_B}}{d s_B}
\end{equation}
with $s_B \in [0,\pi]$, and $\theta_B(s_B) = \theta_A - s_B \, (1 + 2 \, \theta_A/\pi)$. For this part $B$, $| \bm{t_B}(s_B) | \neq 1$.
Part $C$ is a right-handed helix of radius unity, total arclength $L_C=L_A$ and helical angle $\theta_C=-\pi-\theta_A=\pi-\theta_A \, mod \, 2 \pi$:
\begin{equation}
\bm{r_C}= \left(
\begin{array}{c}
- \sin \psi_C(s_C)\\
 \cos \psi_C(s_C) \\
L_A \cos \theta_A + s_C \, \cos \theta_C
\end{array}
\right)
 \mbox{ and } 
\bm{t_C}=\frac {d \bm{r_C}}{d s_C} = \left(
\begin{array}{c}
\sin {\theta_C} \, \cos \psi_C(s_C)\\
\sin \theta_C \, \sin \psi_C(s_C) \\
\cos \theta_C
\end{array}
\right)
\end{equation}
with $s_C \in [0,L_C]$ and $\psi_C(s_C)=\Delta \psi_A - s_C \, \sin \theta_C$. 
We have $| \bm{t_C}(s_C) |=1$ for all $s_C$, and consequently $s_C$ is the 
arclength in part $C$. 
The end loop $D$ closes the curve:
\begin{equation}
\bm{r_D}= \left(
\begin{array}{c}
  \sin \theta_A \sin s_D \cos s_D  \\
  \cos s_D  \\
  \left( \sin \theta_D(s_D) - \sin \theta_A   \right) / (1 + 2 \, \theta_A/\pi) 
\end{array}
\right)
\mbox{ and } \bm{t_D}=\frac {d \bm{r_D}}{d s_D}
\end{equation}
with  $s_D \in [0,\pi]$, and $\theta_D(s_D) = -\pi - \theta_A + s_D \, (1 + 2 \, \theta_A/\pi)$. 
For this part $D$, $| \bm{t_D}(s_D) | \neq 1$.
We call $\Gamma$ the union of the four parts: $\Gamma = \Gamma_A \cup \Gamma_B \cup \Gamma_C \cup \Gamma_D$. The curve has continuous derivatives, and finite jumps in its curvature.
\subsection*{Analytical calculation}

The evaluation of the double integral of Eq.~(\ref{equa:writhe_def}) in the limit $L_A \gg 1$ yields
 \cite{neukirch+heijden:2002,starostin:2002,starostin:2005}:
\begin{equation}
Wr^{CW}(\Gamma) \stackrel{L_A \to \infty }{\longrightarrow} - \frac{L_A}{\pi} \sin \theta_A \cos \theta_A \, .
\label{equa:double_integral_formula}
\end{equation}
The minus sign is due to the fact that the helices $A$ and $C$ are right-handed.

\begin{figure}[htbp]
\begin{center}
\includegraphics[angle=0,width=0.50\linewidth]{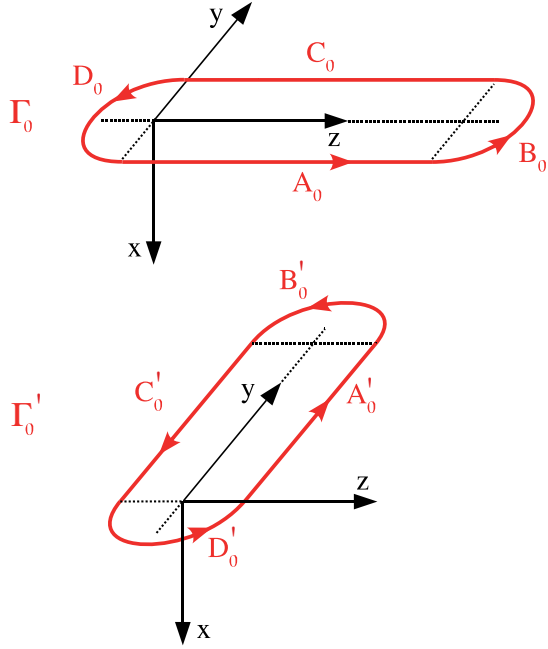}
\caption{Two references curves used to compute $Wr^F$ for the supercoiled plasmid of Fig.~\ref{fig:clamped_ply}.}
\label{fig:two_reference_curves_plasmid}
\end{center}
\end{figure}

We then compute $Wr^F(\Gamma, \Gamma_0)$, given by Eq.~(\ref{equa:fuller_2nd_theo}),
using the reference curve $\Gamma_0$ of Fig.~\ref{fig:two_reference_curves_plasmid}, top.
The continuous deformation between $\Gamma_0$ and $\Gamma$ is then
parametrized by $\theta_{A_\lambda} \in [0,\theta_A]$.
The reference curve of Fig.~\ref{fig:two_reference_curves_plasmid} top comes to mind
naturally and makes calculations easiest, but yields antipodal events
and hence an incorrect result, as we shall see now.
The writhe of the reference curve  $\Gamma_0$ is zero. Consequently we only have to compute the integral $Fu(\Gamma, \Gamma_0)$ of Eq.~(\ref{equa:fuller_2nd_theo})
over the four different parts $A$, $B$, $C$, and $D$.
Clearly when the plasmid is long enough, i.e. when $L_A$ is large, the contribution to $Fu(\Gamma, \Gamma_0)$ of the helical parts $A$ and $C$ become dominant. Indeed their contribution scales with $L_A$ while the contribution of the end loops $B$ and $D$ remains bounded.
Consequently we focus on the contributions of
the helical parts $A$ and $C$, for large values of $L_A$, that is we look at the limit $Fu(\Gamma, \Gamma_0) \to Fu_A(\Gamma, \Gamma_0)+Fu_C(\Gamma, \Gamma_0)$
when $L_A \to \infty$ \cite{crick:1976}.
For the helical part $A$, the corresponding tangent of the reference curve is $\bm{t_{A_0}}=(0,0,1)$. Fuller integral for part $A$ is then:
\begin{equation}
Fu_A= \int_0^{L_A} \frac{d\psi_A}{ds_A} \, (1-\cos\theta_A) ds_A = \sin \theta_A \, (1-\cos \theta_A)  \, L_A \, .
\end{equation}
For the helical part $C$, the corresponding tangent of the reference curve is $\bm{t_{C_0}}=(0,0,-1)$. The Fuller integral for part $C$ is then:
\begin{equation}
Fu_C= \int_0^{L_C} -\frac{d\psi_C}{ds_C} \, (1+\cos{\theta_C}) ds_C
= \sin \theta_C \, (1+\cos \theta_C)  \, L_C
= Fu_A
\end{equation}
Neglecting the end loop contributions we arrive at:
\begin{eqnarray}
Wr^F(\Gamma,\Gamma_0)&=&Wr(\Gamma_0)+\frac{1}{2\pi} Fu(\Gamma, \Gamma_0) \\
&=&\frac{1}{2\pi} \left( Fu_A+Fu_B+Fu_C+Fu_D \right) \\ 
&\simeq& \frac{1}{2\pi} \left(  Fu_A + Fu_C  \right) =\frac{L_A}{\pi} \sin \theta_A (1-\cos \theta_A)
\label{equa:wrong_Fu_classic_reference_curve}
\end{eqnarray}
We first remark that $Wr^F(\Gamma,\Gamma_0)>0$ although $Wr^{CW}(\Gamma) < 0$ (see Eq.~(\ref{equa:double_integral_formula})).
The discrepancy $Wr^F(\Gamma) \neq Wr^{CW}(\Gamma)$ is due to the fact that in the continuous deformation 
$\theta_{A_\lambda} \in [0,\theta_A]$
from the curve $\Gamma_0$ of Fig.~\ref{fig:two_reference_curves_plasmid} top to the
plasmid of Fig.~\ref{fig:clamped_ply}, there are antipodal events.
These events happen each time the middle point of the end loop $B$ is pointing towards $-\bm{e_y}$ and hence become antipodal to the corresponding tangent in $B_0$. 
As $\theta_{A_\lambda}$ is increased, such an event happens once each time $\Delta \psi_{A_\lambda} \DEF L_A \, \sin \theta_{A_\lambda}$ increases by an amount of $2\pi$.
In the entire deformation there will be $\Delta \psi_A / (2\pi)$ antipodal events.
Each event introduces a shift between $Wr^{CW}(\Gamma)$ and $Wr^F(\Gamma,\Gamma_0)$ of $2$, the difference $Wr^F(\Gamma,\Gamma_0)-Wr^{CW}(\Gamma)$ should then be equal to $2 \, \Delta \psi_A / (2\pi) = (L_A/\pi) \, \sin \theta_A$, which can be verified by comparing Eq.~(\ref{equa:double_integral_formula}) to Eq.~(\ref{equa:wrong_Fu_classic_reference_curve}).
A natural way to try to avoid the presence of these antipodal points is to rotate the reference curve, e.g. choose ${\Gamma_0}'$ of Fig.~\ref{fig:two_reference_curves_plasmid} bottom as reference curve.
In this  case $Fu_A=-\cos \theta_A \sin \theta_A \int_0^{\Delta \psi_A} \sin \psi / (1 + \sin \theta_A \sin \psi) \; d\psi$ and $Fu_C=\cos \theta_A \sin \theta_A \int_0^{\Delta \psi_A} \sin \psi / (1 - \sin \theta_A \sin \psi) \; d\psi \; \neq Fu_A$. Here again the sum $Fu_A + Fu_C$ dominates $Fu(\Gamma, \Gamma_0')$.
Integration shows that $Fu_A + Fu_C=2 L_A \sin \theta_A (1-\cos \theta_A) + \eta(L_A,\theta_A)$, with $| \eta(L_A,\theta_A) | \leq 1-\cos \theta_A$, which is not the correct result.
Antipodal events are in fact still present with this rotated reference curve $\Gamma_0'$ and trying another rotation will not help, as explained in Appendix B.

\subsection*{Numerical verification}
%
%

In Fig.~\ref{fig:clamped_ply}, a right-handed circularly closed plasmid is drawn with $\theta_A=\pi/6$, $L_A=5 \pi$.
The viewpoint selected to draw the curve in Fig. \ref{fig:clamped_ply} is such that three (negative) crossings appear. For some other viewpoints, only two crossings appear.
The writhe being the average number of signed crossings one sees from all possible viewpoints,
it is relatively easy to convince oneself that its value for  the curve of Fig. \ref{fig:clamped_ply} lies in between -2 and -3.
A numeric discretization scheme \cite{swigon+al:1998} of the double integral of Eq.~(\ref{equa:writhe_def}) with 600 points yields $Wr^{CW}(\Gamma) \simeq -2.2049$.
A numerical integration of $Fu(\Gamma, \Gamma_0)=Fu_A+Fu_B+Fu_C+Fu_D$
yields $Wr^F(\Gamma,\Gamma_0)\,(=Wr^F(\Gamma,\Gamma_0')) \simeq -0.2050$. The difference between $Wr^{CW}(\Gamma)$ and $Wr^F(\Gamma,\Gamma_0)$ is -2, which indicates that the net number of signed antipodal events is -1, as shown in Appendix B.



 %
 %
 %
 %
 %
 %
 %
 %
 %
 \section{Appendix B : The closed circuit theorem} \label{appendix 2}
 %
 %
 %
 %
 %
 %
 %
 %
\subsection*{Antipodal points and self-crossings}
\begin{figure}[htbp]
\begin{center}
\includegraphics[angle=0,width=0.70\linewidth]{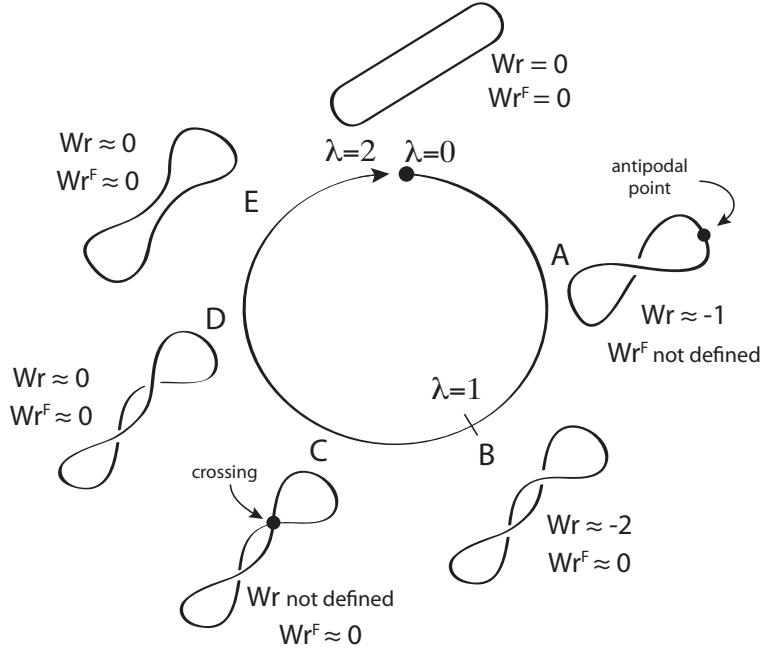}
\caption{A closed circuit with one antipodal point and one self-crossing.}
\label{fig:closed_circuit}
\end{center}
\end{figure}

%
%
In Fig.~\ref{fig:closed_circuit} we show a family of curves along a continuous deformation. The deformation starts and ends with the same curve: we have a closed circuit.
The `stadium' shaped curve $\Gamma_{\lambda=0}$ is continuously deformed through curves $\Gamma_\lambda$, as $\lambda$ increases from $\lambda=0$ to $\lambda=1$, to the supercoiled plasmid $\Gamma_{1}$.
The plasmid is then deformed back to the `stadium' shaped curve, through another path, as $\lambda$ increases from $\lambda=1$ to $\lambda=2$.
The values $\lambda=0$ and $\lambda=2$ refer to the same curve, the reference curve, which plays a special role.
This reference curve is chosen to be fairly simple, in particular $Wr(\Gamma_{0})=0$.
All the curves in the continuous deformation share a parametrization with parameter $s$ (not necessarily the arclength) taking values from $s=0$ to $s=L$.
We recall that an antipodal point is a point on a curve $\Gamma_\lambda$ such that the tangent $\bm{t}_\lambda(s^{\star})$ to $\Gamma_\lambda$ at $s=s^\star$ is aligned and in opposite direction with the tangent $\bm{t}_0(s^{\star})$ to the reference curve $\Gamma_0$, i.e. an antipodal point is such that
$\bm{t}_\lambda(s^{\star}) \cdot \bm{t}_0(s^{\star}) = -1$. An antipodal point is always defined with regard to a certain reference curve.
Along  a continuous deformation there may be $\lambda$ values where the curves $\Gamma_\lambda$ have antipodal points to $\Gamma_0$, and there may be $\lambda$ values where the curves  $\Gamma_\lambda$ self-intersect.
{\em The closed circuit theorem} claims that in any closed circuit the (signed) number of self-intersections is equal to the (signed) number of antipodal points.
For example, in the closed circuit of Fig.~\ref{fig:closed_circuit} there is  one antipodal point and one self-crossing.
This theorem was proposed and proven in \cite{samuel+al:2006} for a particular case of reference curve,  a straight line.
Here we give an outline of a proof which is valid for more general reference curves:~Formula (\ref{equa:writhe_def}) has a $\pm 2$ discontinuity at each self-crossing event encountered in the closed circuit, but no discontinuity during antipodal events. On the contrary formula (\ref{equa:fuller_2nd_theo}) has no discontinuity during self-crossing events but a $\pm 2$ discontinuity at each antipodal event \cite{aldinger:1995}.
Formula (\ref{equa:writhe_def}) and (\ref{equa:fuller_2nd_theo}) by definition agree when applied on the reference curve, and furthermore they may only differ by an even integer: $Wr^{CW}=Wr^{F} mod ~2$.
If we now follow the closed circuit, each formula will have its own discontinuities but both must eventually agree again at the end of the circuit. It follows that the number of discontinuities (respectively events) of one kind must be equal to the number of discontinuities (respectively events) of the other kind.
\subsection*{No parallel between $Wr^{CW}(\Gamma)$ and $Wr^F(\Gamma,\Gamma_0)$}
%
%
%
%
%
The quantity $Wr^{CW}(\Gamma)$ is not defined on curves that are self-intersecting, e.g. curve $C$  in Fig.~\ref{fig:closed_circuit}.
The quantity $Wr^F(\Gamma,\Gamma_0)$ is not defined on curves $\Gamma$ that have an antipodal point to $\Gamma_0$, e.g. curve $A$  in Fig.~\ref{fig:closed_circuit}.
These two properties, once considered with the fact that in any closed circuit the number of configurations with self-crossing is equal to the number of configurations with antipodal points, could lead one to believe that some parallel exists between $Wr^{CW}(\Gamma)$ and $Wr^F(\Gamma,\Gamma_0)$.
In fact in \cite{samuel+al:2006} it is proposed that, in the computation of the mean writhe (or link) of a statistical ensemble of curves, avoiding self-crossings was equivalent to avoiding antipodal points. Moreover it was infered that for such statistical ensemble of curves, Fuller's formula, Eq.~(\ref{equa:fuller_2nd_theo}),
could be used safely provided that curves with antipodal point(s) were discarded when generating the statistical ensemble.
We argue that this is not the case as Fuller's formula does not yield the right result for the (many) curves that have no antipodal point: e.g. curve 
$B$ in Fig.~\ref{fig:closed_circuit}.
Again, Fuller's formula can only be used for a curve $\Gamma_{\bar{\lambda}}$ if one can devise a continuous deformation from $\Gamma_{\bar{\lambda}}$ to $\Gamma_0$ where {\em none} of the curves $\Gamma_\lambda$ with $0< \lambda \leq \bar{\lambda}$ has antipodal points.
For curve $B$  in Fig.~\ref{fig:closed_circuit} this is not the case since the curve  $A$ has an antipodal point.
Accordingly Fuller's formula does not yield the correct value for the writhe of curve $B$, as explicitly computed in Appendix A.
In conclusion we stress that there is no parallel between the two formulas 
$Wr^{CW}(\Gamma)$ and $Wr^F(\Gamma,\Gamma_0)$: on the one hand $Wr^{CW}(\Gamma)$ yields the correct value of the writhe of a curve as soon as the curve does not self-intersect and on the other hand $Wr^F(\Gamma,\Gamma_0)$ often yields an incorrect value for the writhe of a curve even if this curve has no antipodal point.

All what is exposed here for the curves in Fig.~\ref{fig:closed_circuit} is directly applicable to the configurations of Fig.~\ref{fig:antipodal_plectonems} and shows that the writhe (resp. the link) of the configurations with $Z < Z_B$ in Fig.~\ref{fig:diag_z_de_n} is not given by Eq.~(\ref{equa:fuller_2nd_theo}) (resp. Eq.~(\ref{equa:link_phi_psi})).
\subsection*{Impossibility to use the `stadium' shaped curve as reference curve}
%
%
%
%
%
%
Fuller's theorem \cite{fuller:1978} states that as soon as there is a continuous deformation between $\Gamma$ and $\Gamma_0$ which is free of antipodal points and self-crossings (we called `good' such a deformation), 
$Wr^F(\Gamma,\Gamma_0)$ yields the correct value for the writhe
of the closed curve  $\Gamma$.
In Fig.~\ref{fig:closed_circuit} the deformation from $\lambda=0$ to $\lambda=1$ is a `bad' deformation as it contains one antipodal event. We then conclude that $Wr^F(\Gamma_1,\Gamma_0)$ yields an incorrect value.
But one might say that another deformation between $\Gamma_1$ and $\Gamma_0$, with neither self-crossing nor antipodal event, could exist.
We show here that this is not the case. {\em Statement}: no `good' deformation between $\Gamma_1$ and $\Gamma_0$ of Fig.~\ref{fig:closed_circuit} exists. {\em Proof}: assume a `good' deformation exists. Then we can build a closed circuit by joining this proposed deformation and the (reversed) actual deformation $\lambda \in [0,1]$ in Fig.~\ref{fig:closed_circuit}. One would then have a closed circuit with one antipodal point and no self-crossing, in contradiction with the result established in
\cite{samuel+al:2006}.
Hence no such `good' deformation exists.$\square ~$ 
This means that as soon as one `bad' deformation exists between a reference curve $\Gamma_0$ and a curve $\Gamma$ the formula $Wr^F(\Gamma,\Gamma_0)$ does not yield the correct result and no `good' deformation can exist.
A consequence of this is that rotating the reference curve of Appendix A (Fig.~\ref{fig:two_reference_curves_plasmid}, top) to introduce a new, and `good', reference curve (e.g. Fig.~\ref{fig:two_reference_curves_plasmid}, bottom) is hopeless.
To show this we first introduce a `bad' deformation between the rotated reference curve (e.g. Fig.~\ref{fig:two_reference_curves_plasmid}, bottom) and the supercoiled plasmid of Fig.~\ref{fig:clamped_ply}: we untangle the plasmid using a self-crossing, as in Fig.~\ref{fig:closed_circuit}-C, to obtain a `stadium' shaped curve which we subsequently rotate to the (rotated) reference curve.
The existence of this `bad' deformation means that no `good' deformation can exist.

%
%
%
%
%
%
%
%
%


%
%
%
%
%
\bibliographystyle{plain}
\end{document}